# A Model for Predicting Ignition Potential of Complex Fuel in Diurnally Variable Environment

Saurabh Saxena, Ritambhara Dubey, Neda Yaghoobian[*]

Florida State University, FAMU-FSU College of Engineering, Department of Mechanical Engineering, Tallahassee, FL, USA

## Abstract

Fuel ignition potential is one of the primary drivers influencing the extent of damage in wildland and wildland-urban interface fires. Determining fire and ember exposure of fuels that vary spatially and temporally will help to recognize necessary defensive actions and reduce damages. In this paper, the development of a new computational model, Temperature And Moisture Evolution predictor for complex Fuel in Open Environment (TAMEFOE), is presented. TAMEFOE predicts the diurnal temperature and moisture content evolution and vulnerability to flame ignition of objects/fuels with complex shapes or settings and materials under variable environmental conditions. The model is applicable to complex fuel scenarios (e.g., interface or intermix communities) composed of natural and manmade random-shaped objects in open atmosphere under the influence of local weather and diurnal solar radiation. The vulnerability of fuel to ember or fire ignition is determined by predicting the transient temperature and dryness of fuel in connection with the surrounding, local environment, and flame heat if any exists. In this regard, a detailed surface energy balance analysis, coupled with a water budget analysis, is performed in high spatiotemporal resolution. The model performance was validated against several existing analytical and measured data. The discrete, high-resolution surface temperature and moisture content information obtained from the model can also provide unsteady boundary conditions for computational fluid dynamics simulations when coupled physics is desired.

## 1. Introduction

Flame and ember ignition vulnerability of fuels is the major factor in determining the scale of community damages from wildfires. In the recent National Institute of Standards and Technology (NIST) technical note (Maranghides et al., 2021), reporting the Camp Fire progression timeline, fire ignition potential is categorized as the first (among four) primary causes of the extensive devastation. The other three factors were identified as density of vegetative and structural fuels, wind and terrain, and extent/size of fire front reaching the communities. Fuel ignition potential includes the receptivity of fuel to direct flame (through radiation and convection) and ember ignition. According to the NIST technical note, in Camp Fire, fuel receptivity and ignition from embers caused a large number of ignitions within the communities and were reported in several first responder statements. It is stated that "*in Paradise, these ignitions started approximately 30 min to 40 min before the arrival of the fire front and rapidly grew in number when the front reached the community*" (Maranghides et al., 2021).

Readiness and vulnerability of fuels to fire and ember, determined by the fuel temperature and dryness, can vary significantly spatially and temporally, depending on the fuel material, time of the day/year, local weather conditions, terrain and topography, geographical location, and proximity to fire flame if any exists. In the same NIST Camp Fire report, it has been emphasized that the fuel ignition potential can vary on scales less than one fourth of acre. Scientific information for quantifying fire and ember exposure is currently limited and measurements are the recommended methods to be used to quantify these exposures (Maranghides et al., 2021). However, conducting local measurements, especially in transient and unsteady conditions, is limited, costly, and laborious and its feasibility depends on the scale and accessibility of the fuels under consideration.

In this work, we aim to introduce a new physics-based computational tool that can predict the spatiotemporally variable temperature and dryness, and ignition readiness of complex fuels in transient environmental conditions. This is done through detailed coupled surface energy and water balance analyses in high spatiotemporal resolution under diurnally variable environmental conditions. To the best of our knowledge, such capability is currently missing, and the scope and potentials of the existing models serve purposes other than those required for complex community fires. For example, the Ignition Potential Model (Hadjisophocleous & Thomas, 1996), developed by the National Research Council of Canada, is an indoor-environment model that uses statistical values to estimate the relative likelihood of ignition in a specific building. The model calculates an ignition potential factor based on the building characteristics related to ignition. This factor indicates the likelihood of fire ignition in the building in comparison to the statistical average for buildings of the same occupancy group. There are models that derive ignition probability from live fuel moisture content (LFMC) data (e.g., Chuvieco et al. 2004; Dennison et al. 2008; Dennison & Moritz 2009, Chuvieco et al. 2009; Jurdao et al. 2012). These models use experimentally found live fuel biophysical parameters or statistical relations between fuel moisture content and historic fire occurrence to define a critical LFMC below which the ignition potential increases significantly. These empirically based models are limited and difficult to be extended to other areas (Jurdao et al. 2012).

The introduced TAMEFOE model provides a framework for improving fire risk assessments by predicting where and when a fire is more prone to ignite, where it may cause more negative impacts, or when fuels in complex scenarios reach the ignition point if they are exposed to fire heat at different times of the day. Such simulation of transient surface temperature and moisture requires detailed energy and water balance analyses in high spatial and temporal resolutions, which is challenging, especially for random-shaped multi-material objects in complex environments. The majority of the existing computational models that perform surface energy balance analysis (to find surface temperatures) are introduced in the field of urban microclimatology. In these models, the objects under investigation are mainly simple-shaped surface-mounted cubes or two-dimensional rectangles, representing buildings. Some examples are the models by Arnfield (1990), Wu (1995, URBAN 4), Mills (1997), Kanda et al. (2005, SUMM), Krayenhoff & Voogt (2007, TUF-3D), Yaghoobian & Kleissl (2012, TUF-IOBES), Asawa et al. (2008), Hénon et al. (2012, SOLENE), Liu et al. (2012), Yang & Li (2013, MUST), and Lee & Lee (2020, MUSE). In addition to handling simple-shaped objects in the absence of moisture, in some of these models, the imposed environmental conditions are steady and thus, do not represent the transient forcing due to the variable radiative conditions (e.g., Arnfield 1990), the incident radiation is prescribed based on empirical relationships with constant values of solar intensity (e.g. Wu 1995; Mills 1997; Yang & Li 2013), the whole urban region is made of one uniform material (e.g., Arnfield 1990; Kanda et al. 2005), or only a bulk-value surface temperature is simulated for an entire building façade, ignoring sub-facet scale temperatures and the transient shadow effects (e.g., Mills 1997; Kanda et al. 2005). There have been a few conduction-only-based studies that handle more complex non-cubic objects (Baek et al. 1998; Kim et al. 2001; Byun et al. 2003; Zabihi et al. 2017). However, these simple models do not fall in the category of transient-environment models that perform surface energy balance analysis as they only provide the internal-body temperature of the objects based on the conduction process as a result of prescribed fixed surface temperatures.

This work introduces the TAMEFOE model theory and demonstrates that the model performs well relative to available observations. Description of the model and model validation studies are, respectively, provided in Sect. 2 and 3. In Sect. 4, an example of the model application is presented.

Sect. 5 details a study of the model sensitivity to different parameters, followed by the conclusion in Sect. 6.

## 2. Model Design
### 2.1 Handling Complex Geometry and the Surface Energy Budget Equation

In TAMEFOE, 3D complex random-shaped objects or environments are modeled by defining triangular mesh patches over the surfaces. Each surface patch is defined by its vertex coordinates that form the face of the patch and an outward vector normal to the face. Individual patch orientation is specified by its normal with no restrictions to the direction, which permits the model to handle any complex surface. Each patch is in instantaneous energy exchange with its immediate environment and surrounding patches/objects. The modeled objects can be surface-mounted or non-surface-mounted. The energy balance calculations are performed over each individual triangular patch, providing a high-resolution prediction of surface temperature evolutions. The energy balance calculation is done by considering the geographical location (allowing for the accurate calculation of diurnal solar position and flux), local (minute or hourly) weather conditions provided by the typical meteorological year (TMY) weather data files (NSRDB), and characteristics of the surface and substrate materials of each patch. Characteristics of the materials are defined by their physical, thermal, and radiative properties. TMY files provide meteorological data for typical days of a typical year in different geographical locations. The fundamental formula for the energy equilibrium at a triangular patch $i$ (Eq. (1)) includes: 1) net radiative heat exchanges ($R_{net,i}$ (W)) comprising of net shortwave ($R_{SW,net,i}$ (W)) and net longwave ($R_{LW,net,i}$ (W)) radiations absorbed over the patch, 2) convective heat exchanges at the patch surface including sensible ($Q_{H,i}$ (W)) and latent ($Q_{E,i}$ (W)) heat, and 3) conduction of heat ($Q_{G,i}$ (W)) normal to the patch material layers. In the presence of a heat source a heat input ($Q_{F,i}$ (W)) is also considered. Each of these terms is described in the following subsections.

$$R_{net,i} = Q_{G,i} + Q_{H,i} + Q_{E,i} + Q_{F,i} \tag{1}$$

Equation (1) is a fourth-order equation that is solved for the surface temperature of patch $i$ ($T(i,t)$ (K)) at each time step ($t$) using the Newton-Raphson method until the patch surface temperature varies less than 0.001K between iterations. For simplicity, $T(i,t)$ will be represented as $T_i$ in the following. To calculate the moisture content within the patch material, a separate energy and water budget is simultaneously solved within each material layer using the Newton-Raphson method and then coupled with Eq. (1) via the conduction and latent heat terms.

**2.2 Net Radiative Heat Exchange Rate**

The net radiative heat exchanges represent a combination of shortwave and longwave radiations (Eq. (2)) received over a surface and reflected/emitted by it, each of which is detailed below.

$$R_{net,i} = R_{SW,net,i} + R_{LW,net,i} \qquad (2)$$

*2.2.1 Net Shortwave Radiation*

The quantity of shortwave radiation that is received by each patch is a function of the sun's position, the orientation of the patch with respect to the sun, the presence of any obstructions between the sun and the patch, and the flux of direct solar radiation and diffuse shortwave radiation from the environment and other patches. The obstructions, if any, determine the shaded regions, which are established using the shadow model described in Sect. 2.2.4. The net shortwave radiation received over patch $i$ is, therefore, calculated using:

$$R_{SW,net,i} = R_{SW,i} \downarrow - R_{SW,i} \uparrow \qquad (3).$$

$R_{SW,i} \downarrow$ (W) is the total downwelling shortwave radiation received by the patch and $R_{SW,i} \uparrow$ (W) is the total upwelling shortwave radiation reflected by the patch, defined, respectively, by Eq. (4) and (5):

$$R_{SW,i} \downarrow = R_{SW,direct,i} + R_{SW,diffuse,i} + \sum_{m=1}^{threshold} \sum_{j=1}^{npatch} F_{ji}\left(R_{SW,j}^{m-1} \uparrow\right) \qquad (4),$$

$$R_{SW,i} \uparrow = a_i \, R_{SW,i} \downarrow \qquad (5).$$

In Eq. (4), $R_{SW,direct,i}$ (W) is the incident direct solar radiation and $R_{SW,diffuse,i}$ (W) is the incident diffuse shortwave radiation (defined, respectively, in Eq. (9) and (10)), and the last term is the summation of the shortwave radiation reflected from other surrounding patches received over patch $i$. $npatch$ is the total number of patches in the domain. Each patch experiences multiple reflections with $m$ being the reflection number. Reflections are computed until the remaining unabsorbed radiation is below a user-defined threshold or until a minimum user-defined number of reflections is achieved (represented in the term $threshold$). $a_i$ in Eq. (5) is the patch surface albedo, which depends on the patch surface material and determines the amount of shortwave radiation reflected by the surface back to the environment.

Equation (6) details the total upwelling shortwave reflection from patch $i$. $R_{SW,i}\uparrow$ includes the initial reflected shortwave radiation ($R_{SW,i}^0\uparrow$ (W)) and other subsequent reflections ($R_{SW,i}^m\uparrow$ (W)) from the patch. The initial reflection ($R_{SW,i}^0\uparrow$ (W); Eq. (7)) is the sum of the reflected fractions of the incident direct solar and diffuse shortwave radiations. Subsequent reflections ($R_{SW,i}^m\uparrow$ (W); Eq. (8)) are fractions of the previous reflections from other surrounding patches $j$, allowing for the consideration of multiple reflections in the model.

$$R_{SW,i}\uparrow = R_{SW,i}^0\uparrow + \sum_{m=1}^{threshold}\left(R_{SW,i}^m\uparrow\right) \quad (6),$$

$$R_{SW,i}^0\uparrow = a_i R_{SW,direct,i} + a_i R_{SW,diffuse,i} \quad (7),$$

$$R_{SW,i}^m\uparrow = a_i \sum_{j=1}^{npatch} F_{ji}\left(R_{SW,j}^{m-1}\uparrow\right) \quad (8).$$

The incident direct solar and diffuse shortwave radiations in Eq. (4) are found from:

$$R_{SW,direct,i} = \boldsymbol{n_i}\cdot\boldsymbol{R}A_i \quad (9),$$

$$R_{SW,diffuse,i} = \left(1 - \sum_{j=1}^{npatch} F_{ij}\right)\cdot I_{diffuse}A_i \quad (10).$$

In Eq. (9), $\boldsymbol{R}\;(=|I_{direct}|\boldsymbol{r_{unit}})$ is the solar radiation vector with $\boldsymbol{r_{unit}}$ being the solar unit vector. $I_{direct}$ (W m$^{-2}$) is the flux of incident direct solar radiation at a particular time for a particular geographical location that is obtained from available weather data files. $A_i$ (m$^2$) is the area of the patch and $\boldsymbol{n_i}$ is the area normal outward unit vector. In Eq. (10), $I_{diffuse}$ (W m$^{-2}$) is the flux of incident diffuse solar radiation of the location attained from available weather data files, $F_{ij}$ is the view factor of patch $i$ with respect to patch $j$ (explained in Sect. 2.2.3), and thus correspondingly $\left(1-\sum_{j=1}^{npatch} F_{ij}\right)$ is the sky view factor of patch $i$.

To calculate the solar unit vector ($\boldsymbol{r_{unit}}$), we first determine the solar position (i.e., solar azimuth ($\emptyset$) and solar zenith ($Z$) angles) following the National Oceanic and Atmospheric Administration (NOAA) solar position calculator (NOAA 2018), for which the latitude, longitude, date, and time zone of the location under consideration is required. The implementation of the NOAA solar position equations in this model has been verified for two locations (Vancouver,

Canada and Guerville, France) with excellent accuracy. The solar unit vector is then estimated from Eq. (11):

$$\boldsymbol{r_{unit}} = cos(\emptyset)\ sin(Z)\ \boldsymbol{e_x} - sin(\emptyset)\ sin(Z)\ \boldsymbol{e_y} + cos(Z)\ \boldsymbol{e_z} \qquad (11),$$

in which $\boldsymbol{e_x}$, $\boldsymbol{e_y}$, and $\boldsymbol{e_z}$ are unit vectors pointing towards the geographical north, west, and vertical directions, respectively.

*2.2.2 Net Longwave Radiation*

In a similar manner to shortwave radiation, the rate of longwave radiative heat received over a patch surface is calculated using the following equations:

$$R_{LW,net,i} = R_{LW,i} \downarrow - R_{LW,i} \uparrow \qquad (12),$$

$$R_{LW,i} \downarrow = R_{LW,sky,i} + \sum_{m=1}^{threshold} \sum_{j=1}^{npatch} F_{ji}\left(R_{LW,j}^{m-1} \uparrow\right) \qquad (13),$$

$$R_{LW,sky,i} = \left(1 - \sum_{j=1}^{npatch} F_{ij}\right) \sigma T_{sky}^4 A_i \qquad (14).$$

Here, $R_{LW,i} \downarrow$ (W) is the total downwelling longwave radiation received by patch $i$ and $R_{LW,i} \uparrow$ (W) is the total upwelling longwave radiation emitted by it. In Eq. (13), $R_{LW,sky,i}$ (W) is the incident longwave radiation from the sky and the last term is the summation of the longwave radiation reflected/emitted from all other surrounding patches received over patch $i$. $T_{sky}$ (K) is the measured air temperature (obtained from weather data files) and $\sigma = 5.67 \times 10^{-8}$ (kg s$^{-3}$ K$^{-4}$) is the Stefan-Boltzmann constant.

The total upwelling longwave radiation from patch $i$ ($R_{LW,i} \uparrow$ (W), see Eq. (15)) is divided into the emitted longwave radiation by the patch due to its surface temperature ($R_{LW,i}^0 \uparrow$ (W), Eq. (16)), the first reflected longwave radiation from the patch ($R_{LW,i}^1 \uparrow$ (W), Eq. (17)), and other subsequent reflections from other patches received over patch $i$ ($R_{LW,i}^m \uparrow$ (W), Eq. (18)). The first reflection ($R_{LW,i}^1 \uparrow$ (W)) is the sum of the reflected fraction of the sky longwave radiation and the reflected fraction of the emitted radiation from all other patches that is incident on patch $i$.

$$R_{LW,i} \uparrow = R_{LW,i}^0 \uparrow + R_{LW,i}^1 \uparrow + \sum_{m=2}^{threshold} \left(R_{LW,i}^m \uparrow\right) \qquad (15),$$

$$R_{LW,i}^0 \uparrow = \varepsilon_i \sigma T_i^4 A_i \qquad (16),$$

$$R_{LW,i}^1 \uparrow = \rho_i R_{LW,sky,i} + \rho_i \sum_{j=1}^{npatch} F_{ji}\, \varepsilon_j \sigma T_j^4 A_j \tag{17},$$

$$R_{LW,i}^m \uparrow = \rho_i \sum_{j=1}^{npatch} F_{ji} \left(R_{LW,j}^{m-1} \uparrow\right) \tag{18}.$$

In the above equations, $T_i$ (K) is the surface temperature of patch $i$ (to be found), $\varepsilon_i$ is the emissivity of the patch and it is equal to the surface absorptivity, and $\rho_i$ is the longwave reflectivity of patch $i$, assuming the transmissivity of the patch is zero.

*2.2.3 View Factor*

The view factor ($F_{ij}$) of patch $i$ represents the fraction of radiation leaving the patch surface and reaching another patch $j$. This factor depends on the visibility of each pair of patches ($i$, $j$) with respect to each other, their sizes, distance, and relative orientation. For a pair of patches ($i$, $j$), the dot product between the outward normal vector of each patch ($\boldsymbol{n_i}$ or $\boldsymbol{n_j}$) and the vector joining centroids of the two patches ($\boldsymbol{c_j c_i}$ or $\boldsymbol{c_i c_j}$), i.e., $\boldsymbol{n_j} . \boldsymbol{c_j c_i}$ and $\boldsymbol{n_i} . \boldsymbol{c_i c_j}$, determines the visibility of each of the patches with respect to each other. If these dot products are positive, the pair of patches can see each other.

The most commonly used methods to calculate view factors between surfaces are the Monte Carlo method (e.g., in Hoff & Janni, 1989; Howell, 1998; Mirhosseini & Saboonchi, 2011) and the contour integral method (e.g., in Sparrow, 1963; Rao & Sastri, 1996; Mazumder & Ravishankar, 2012). The Monte Carlo method computes view factors as the ratio of the rays incident on a surface patch to the total rays emitted from the other surface based on the radiative emission from that surface. However, since it requires several ray-tracing calculations, it is computationally expensive for 3D surface geometries with high-resolution meshes. Thus, in this work, we employ a contour integral method using vector parametric representation as proposed in Mazumder & Ravishankar (2012). The details of the model are briefly presented in Appendix A.

*2.2.4 Shadow Model*

The presence of any obstruction between the sun and any triangular patch $i$ is determined using a ray-tracing method. An obstruction creates a shadow region for patch $i$, which is estimated by assessing a direct line of sight between the sun and the patch and locating the intersection of this line and the plane of potential obstructing patches ($j$). This direct line of sight is represented using a 3D equation of a line in vector format, which is calculated using the position vector of the centroid of patch $i$ ($c_i$) and the solar unit vector ($r_{unit}$). Here, the position vector defines the position of any point with respect to a fixed origin (0,0,0). Thus, any point on this line can be found using the centroid of patch $i$, which lies on the line and a multiple of the unit vector $r_{unit}$ that is parallel to this line as shown below:

$$x = c_i + l_p\, r_{unit} \qquad (19a),$$

$$(x - c_{ix}) = l_p\, r_{unit,x};\ (y - c_{iy}) = l_p\, r_{unit,y};\ (z - c_{iz}) = l_p\, r_{unit,z} \qquad (19b).$$

In Eq. (19), $c_{ix}$, $c_{iy}$, $c_{iz}$ are the $x$, $y$, $z$ components of the position vector $c_i$, $r_{unit,x}$, $r_{unit,y}$, $r_{unit,z}$ are the $x$, $y$, $z$ components of the solar unit vector, and $l_p$ is the parameter describing any point on the line. To estimate whether there is any obstruction over this direct line of sight between patch $i$ and the sun, the 3D equation of each plane containing other patches ($j$) is found using:

$$(x - c_j) \cdot n_j = 0 \qquad (20).$$

Equation (20) describes the vector equation for a plane with a given position vector and a normal. Here, $n_j$ is the outward unit normal of patch $j$ and $c_j$ is the position vector of the centroid of patch $j$. Subsequently, the position vector of the intersection point ($c_{inter}$) between the plane through patch $j$ and the direct line of sight between patch $i$ and the sun is determined. This intersection point lies on both the plane containing patch $j$ (Eq. (20)) and the direct line of sight (Eq. (19a)). Thus, the two equations (Eq. (19a) and Eq. (20)) are solved simultaneously and the parameter $l_p$ is evaluated to find the position vector of the intersection point ($c_{inter}$):

$$l_p = -n_j \cdot (c_i - c_j) / (n_j \cdot r_{unit}) \qquad (21),$$

$$c_{inter} = c_i + l_p\, r_{unit} \qquad (22).$$

If this intersection point falls within the triangular patch $j$, and the angle ($\theta$) between the normal of patch $i$ ($n_i$) and the direct line of sight is between 0 and 90 degrees, then patch $i$ is deemed to be shaded. Figure 1 shows the diagrammatic representation and the flow chart of the shadow model. A similar ray-tracing algorithm is used to find obstructions between any two patches.

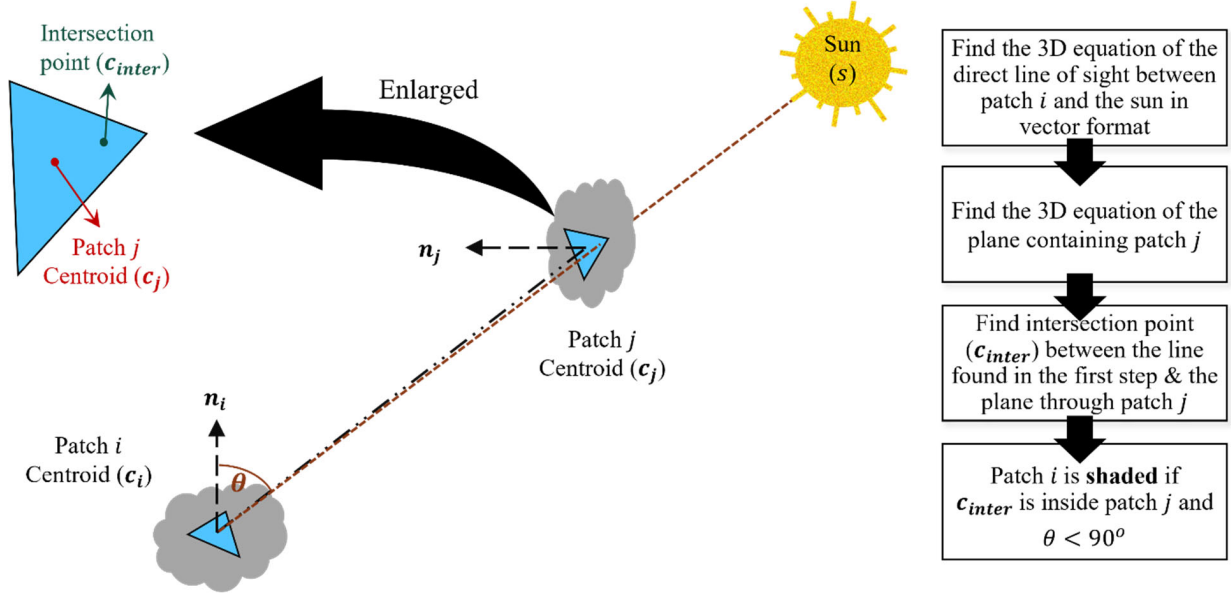

Figure 1: Diagrammatic representation and flow chart of the shadow model, showing patch $i$ to be shaded due to obstruction by patch $j$.

### 2.3 Convective Heat Exchange Rate

The rate of convective heat exchanges at each patch surface is a combination of sensible ($Q_{H,i}$ (W)) and latent ($Q_{E,i}$ (W)) heat exchange rates. In the presence of moisture, the cooling effects of the latent heat on the patch surface temperature in the energy balance equation are modeled using a moisture model (described in Sect. 2.5). The sensible convective heat exchange occurring at the patch surface is computed based on:

$$Q_{H,i} = h_i\,(T_i - T_{air})\,A_i \qquad (23),$$

in which $T_{air}$ (K) is the ambient air temperature, $T_i$ (K) is the surface temperature of patch $i$ (to be found), and $h_i$ (W m$^{-2}$ K$^{-1}$) is the sensible heat transfer coefficient of the patch. $h_i$ is approximated either (a) using the bulk (effective) wind speed and air temperature values at the object location, or (b) from the velocity and air temperature local to each individual patch of the object. The former is applied when only bulk values of the ambient wind and air temperature are available, and the latter is applied when high-resolution flow and air temperature data are available, for example, from a computational fluid dynamics (CFD) simulation. For method (a), the effective wind speed and air temperature data at the object location are computed from the reference values obtained from the local weather data files. The reference wind and temperature information in weather files are provided at 10 m and 2 m heights, respectively. Therefore, they are adjusted for the height of each patch according to the log law. The convective heat transfer coefficient ($h_i$) is then estimated using one of the five methods (Nusselt Jurgess, McAdams, the Simple Combined method, TARP, and DOE-2 formulation) included in the model. The details of these methods are provided in Appendix C. In method (b), the same procedure is used for calculating $h_i$ at each patch surface but with velocities and air temperatures local to each patch.

## 2.4 Conductive Heat Exchange Rate

To capture the evolution of temperature within the object body, one-dimensional (1D) conduction within the multi-layer materials of each patch's substrate is considered. Two methods are embedded in the model for the calculation of conduction: a finite difference method (FDM) that uses the Crank-Nicholson time-stepping scheme and an analytical-based Z-transform method that uses the Conduction Transfer Functions (CTFs). The Z-transform method is based on a response factor approach (Stephenson & Mitalas 1967) in which the heat flux at the surface of a material element is related to an infinite series of temperature histories at both the inner and outer extremes of the material layers. The basic form of this approach is:

$$q"(t) = \sum_{cc=0}^{\infty} \left( XX_{cc} T_i^{t-cc\Delta t} + YY_{cc} T_{in}^{t-cc\Delta t} \right) \qquad (24),$$

where $q"(t)$ (W m$^{-2}$) is the heat flux at the surface of an element at the current time $t$ (s), $XX$ (W m$^{-2}$ K$^{-1}$) and $YY$ (W m$^{-2}$ K$^{-1}$) are response factors, $T$ (K) is the temperature with subscripts $i$ and $in$ indicating the outer and inner extremities of the material layer, $\Delta t$ (s) is the time step size, and $cc$ is the number of time steps prior to the current time. Due to the infinite number of response factors required for an exact solution of Eq. (24), Stephenson & Mitalas (1971) further improved the approach by applying the Z-transform theory to the transient heat conduction equation. The improved method forms a conduction transfer solution that reduces the infinite series of response factors by introducing flux history coefficients. This method is computationally efficient and therefore, saves time and computational memory (Stephenson & Mitalas 1971). The conduction transfer solution uses constant CTFs to relate the heat flux at the surface to the current and previous temperatures at the interior and exterior extremities, and the previous heat fluxes. The basic form of the improved Z-Transform equation is:

$$q"_{surface,t} = X_0 T_i^t - Y_0 T_{in}^t + \sum_{cc=1}^{nx} X_{cc} T_i^{t-cc\Delta t} - \sum_{cc=1}^{ny} Y_{cc} T_{in}^{t-cc\Delta t}$$
$$+ \sum_{cc=1}^{nq} \Phi_{cc} q"_{surface,t-cc\Delta t} \tag{25},$$

where, $q"_{surface,t}$ (W m$^{-2}$) is the surface conductive heat flux at the current time step $t$ (s). In this equation, $X$ (W m$^{-2}$ K$^{-1}$), $Y$ (W m$^{-2}$ K$^{-1}$), and $\Phi$ are the constant CTFs for the element's exterior extreme, interior extreme, and surface flux with subscript 0 indicating the current time step. $q"_{surface,t-cc\Delta t}$ (W m$^{-2}$) is the heat flux at the element's outside surface at $cc$ time steps prior to the current time $t$ (s), and $nx$, $ny$, and $nq$ are the maximum number of CTF terms. The CTF coefficients are determined only once for each construction type and are computed using a state-space method (Ceylan & Mayres 1980; Seem 1987), where the maximum number of CTF terms is limited to 19. $q"_{surface,t}$ is then used to find the rate of conductive heat ($Q_{G,i}$ (W)) for each patch in Eq. (1) using the patch surface area ($A_i$ (m$^2$)).

As Eq. (25) shows, the Z-transform method provides a single, linear equation with pre-calculated CFT constants and does not require information from within the material. Therefore, it provides an efficient method for the calculation of conduction and requires a limited space for data storage. As a result, this method significantly reduces the overall computational time in comparison with the FDM.

The FDM is used when information on the temperature evolution within the object material is needed. In the FDM, the conduction into the multiple substrate layers of each surface patch is simulated according to:

$$T_L^{t+1} - T_L^t = \frac{\Delta t}{\rho_L C_L \Delta z_L} \left[ \frac{1}{2} \left( G_{L-1,L}^{t+1} - G_{L,L+1}^{t+1} \right) + \frac{1}{2} \left( G_{L-1,L}^t - G_{L,L+1}^t \right) \right] \tag{26a},$$

$$G_{L-1,L}^t = k_{L-1,L} \frac{(T_{L-1}^t - T_L^t)}{\frac{1}{2} (\Delta z_{L-1} + \Delta z_L)} \tag{26b}.$$

Here, $T_L$ (K), $\rho_L$ (kg m$^{-3}$), $\Delta z_L$ (m), and $C_L$ (J kg$^{-1}$ K$^{-1}$) are, respectively, the temperature, density, thickness, and specific heat capacity of material layer $L$. $G_{L-1,L}^t$ (W m$^{-2}$) is the conductive heat flux between layers $L-1$ and $L$ at timestep $t$, where $k_{L-1,L}$ (W m$^{-1}$ K$^{-1}$) is the conductivity between layers $L-1$ and $L$.

Externally, conduction, in the Z-transform method, is bounded by the patch surface temperature. For the FDM, either the surface energy exchanges that replace the $G_{L-1,L}$ term with $G_{surface,1}$ (Eq. (27a) in Eq. (26a) or following Krayenhoff & Voogt (2007), a conduction flux density condition (Eq. 27b) that is a modified form of Eq. (26a) can be employed to find the temperature of the first layer under the patch surface ($T_1$).

$$G_{surface,1} = R_{net,i} - Q_{H,i} - Q_{E,i} - Q_{F,i} \qquad (27a),$$

$$T_1^{t+1} - T_1^t = \frac{\Delta t}{\rho_1 C_1 \Delta z_1}\left[G_{surface,1}^{t+1} - \frac{1}{2}G_{1,2}^{t+1} - \frac{1}{2}G_{1,2}^t\right] \qquad (27b).$$

Temperatures of all material layers ($T_L$ (K), $L$ from 1 to the deepest layer) are solved at each time step for each patch. Then, the patch surface temperature ($T_i^{t+1}$) at the next time step is solved using the energy balance equation (Eq. (1)), where the conduction term for the FDM is:

$$Q_{G,i} = -k_1 A_i \frac{(T_i^{t+1} - T_1^t)}{0.5 \Delta z_1} \qquad (28).$$

Internally, conduction, in both methods, is bounded by the internal body temperature. The inner extreme of the patch substrate is bounded either by an internal object temperature for a non-ground patch (defined as patches that are not attached to the ground level) or by a deep underground temperature for a ground-level patch. The internal object temperature can be prescribed or can be estimated based on the average surface temperature of the object over some previous timesteps. Following Hillel (1982), the deep underground temperature, $T_{deep}$ (K), at all depths is assumed to vary sinusoidally around an average value, where the amplitude of the fluctuation dampens with the depth below the ground surface based on the thermal properties of the underground material. This estimation reflects the periodic diurnal and seasonal interactions of the soil and atmosphere. $T_{deep}$ at time $t$ (s) and depth $z$ (m) is thus estimated based on:

$$T_{deep}(z,t) = T_{avg} + A_o e^{-\frac{z}{D}} \sin\left(\frac{2\pi t}{P} - \frac{z}{D}\right) \qquad (29),$$

where, $T_{avg}$ (K) is the average surface temperature over a time period $P$ (s), $A_o$ (K) is the amplitude of the surface temperature over the same period, $D = \sqrt{kP/(\rho C_p \pi)}$ (m) is the thermal dampening depth related to the thermal properties of the ground and temperature fluctuations. $k$ (W m$^{-1}$ K$^{-1}$), $C_p$ (J kg$^{-1}$ K$^{-1}$), and $\rho$ (kg m$^{-3}$) are the conductivity, specific heat capacity, and density of the soil/ground. The maximum depth for the calculation of $T_{deep}$ (K) is limited to $z = 3D$ since at this depth the soil temperature amplitude is small (5% or less) compared to the amplitude of the surface temperature ($A_o$) (Hillel, 1982). For the FDM, an additional no heat loss internal boundary condition can be specified.

In general, FDM does not have a geometric or temporal limitation, but it requires storage for each material layer and thus, is computationally costly in comparison to the Z-transform method. On the other hand, the Z-transform method is known to cause stability problems for very short time steps and for when thick materials are under investigation. For this reason, both conduction methods have been implemented in the model and depending on the complexity of the object, time steps, and material layer thicknesses an appropriate method is used.

**2.5 Moisture Model**

A physics-based moisture model, based on Matthews (2006), is implemented and can be utilized in place of any of the conduction modules when the calculation of moisture content, in addition to temperature, within objects/fuels is desired. The model considers all physical parameters important in energy and water budget analyses of the object/fuel solid, intercepted precipitation, and air spaces in the porous material to estimate the internal body temperature and moisture content using FDM.

Assuming horizontal homogeneity, the energy and water budgets of each of the three mediums are estimated in 1D at $N_f$ equally spaced sublayers within the object/fuel. Thus, heat and water budget equations (Eq. 30-32 and 34-36, respectively) for six quantities are analyzed to estimate fuel temperature $T_f$ (K), temperature of liquid water on the solid fuel surface $T_w$ (K), temperature of the internal air spaces $T_{int\_a}$ (K), moisture content $m_f$ (kg water per kg of dry fuel), amount of liquid water $l_w$ (kg of water per m³ of fuel layer), and specific humidity $q_{int\_a}$ (kg of water vapor per kg of air) of each layer under the surface:

$$C_{h,f} \frac{\partial T_f}{\partial t} = \frac{1}{V_f} \left( \frac{\partial R_{fnet}}{\partial z} + \frac{\partial H_{C,f}}{\partial z} \right) - \mu_{f,int\_a} H_{f,int\_a} - \mu_{f,int\_a} \lambda E_{f,int\_a} - \mu_{f,w} H_{f,w} \quad (30),$$

$$C_{h,w} \frac{\partial T_w}{\partial t} = -\mu_{w,int\_a} H_{w,int\_a} - \mu_{w,int\_a} \lambda E_{w,int\_a} + \mu_{w,f} H_{f,w} \quad (31),$$

$$\rho_{int\_a} C_{p,int\_a} \frac{\partial T_{int\_a}}{\partial t} = \frac{1}{V_{int\_a}} \frac{\partial H_{int\_a,T,}}{\partial z} + \mu_{int\_a,f} H_{f,int\_a} + \mu_{int_a,w} H_{w,int\_a} \quad (32).$$

In equations 30–32, subscripts represent medium type namely, the solid fuel ($f$), liquid water ($w$), and internal air ($int\_a$). Equation (30) is the energy budget for the solid fuel medium at each fuel layer. The individual terms in this equation are net radiation ($R_{fnet}$ (W m$^{-2}$)), conduction ($H_{C,f}$ (W m$^{-2}$)), sensible heat flux to internal air ($H_{f,int\_a}$ (W m$^{-2}$)), latent heat flux due to water vapor flux to the internal air ($E_{f,int\_a}$ (kg m$^{-2}$ s$^{-1}$)), and heat transfer between the solid fuel and surface water ($H_{f,w}$ (W m$^{-2}$)), respectively. $C_{h,f}$ (J K$^{-1}$ m$^{-3}$)) is the volumetric heat capacity of the solid fuel, $V_f$ (m$^3$ m$^{-3}$) is the volume of solid fuel per cubic meter of the layer, $\lambda$ (J kg$^{-1}$) is the latent heat of vaporization of water, $\mu$ (m$^{-1}$) is the surface area to volume ratio of the mediums given in the subscripts, $z$ (m) corresponds to the depth of the fuel layer, and $t$ (s) is the time. In this equation, $R_{fnet}$ is the net radiative heat exchanges at the surface of patch $i$, (i.e., $R_{net,i}$ (W)), per the area of the patch. Similar to Eq. (30), Eq. (31) is the energy budget for the liquid water within each layer and the individual terms represent sensible heat flux to the air ($H_{w,int\_a}$ (W m$^{-2}$)), latent heat flux due to evaporation ($E_{w,int\_a}$ (kg m$^{-2}$ s$^{-1}$)), and heat transfer between solid fuel and surface water ($H_{f,w}$ (W m$^{-2}$)), respectively. $C_{h,w}$ (J K$^{-1}$ m$^{-3}$) is the volumetric heat capacity of the liquid water. Lastly, Eq. (32) is the energy budget for the air spaces within each layer with the individual terms representing the heat exchange due to vertical mixing ($H_{int\_a,T}$, (W m$^{-2}$)), sensible heat flux from the solid fuel ($H_{f,int\_a}$ (W m$^{-2}$)), and sensible heat flux to the surface water ($H_{w,int\_a}$ (W m$^{-2}$)). $V_{int\_a}$ (m$^3$ m$^{-3}$) is the volume of solid fuel per cubic meter of the layer, $\rho_{int\_a}$ (kg m$^{-3}$) and $C_{p,int\_a}$ (J kg$^{-1}$ K$^{-1}$) are the air density and the specific heat capacity of air, respectively.

The water or moisture budget for each medium at each fuel layer is calculated using:

$$\rho_{int\_a}\frac{\partial m_f}{\partial t} = -\mu_{f,int\_a}E_{f,int\_a} - \mu_{f,w}E_{f,w} \tag{33},$$

$$\frac{\partial l_w}{\partial t} = \frac{\partial D_r}{\partial z} + \mu_{w,f}E_{f,w} - \mu_{w,int\_a}E_{w,int\_a} \tag{34},$$

$$\rho_{int\_a}\frac{\partial q_{int\_a}}{\partial t} = \frac{1}{V_{int\_a}}\frac{\partial E_{T,int\_a}}{\partial z} + \mu_{int\_a,f}E_{f,int\_a} + \mu_{int\_a,w}E_{w,int\_a} \tag{35}.$$

In these equations, the subscripts represent the medium types as in Eqs. (30-32). Equation (33) is the moisture budget for the solid fuel medium within each layer and the individual terms in the equation are water vapor flux from the litter to the air ($E_{f,int\_a}$ (kg m$^{-2}$ s$^{-1}$)) and liquid water flux from the solid fuel to the surface water ($E_{f,w}$ (kg m$^{-2}$ s$^{-1}$). Similarly, Eq. (34) is the liquid water budget for each layer and the individual terms represent drainage flux ($D_r$ (kg m$^{-2}$ s$^{-1}$)), solid fuel water flux from surface water to solid fuel ($E_{f,w}$ (kg m$^{-2}$ s$^{-1}$)), and evaporation ($E_{w,int\_a}$ (kg m$^{-2}$ s$^{-1}$)), respectively. Equation (35) is the specific humidity budget for the air spaces within each layer with the individual terms representing the heat exchange due to vertical mixing ($E_{int\_a,T}$, (kg m$^{-2}$ s$^{-1}$)), water vapor flux from the solid fuel to air ($E_{f,int\_a}$ (kg m$^{-2}$ s$^{-1}$)), and evaporation from the surface water ($E_{w,int\_a}$ (kg m$^{-2}$ s$^{-1}$)). Through the $E_{f,int\_a}$ and $E_{w,int\_a}$ terms in the above equations, the latent heat ($Q_{E,i}$ (W)) in Eq. (1) is taken care of automatically within the model by calculating the water vapor flux exchange between the fuel and the ambient air at the fuel boundary at each timestep. More details of the model can be found in Matthews (2006).

Equations 30-32 and 33-35 are solved for each fuel layer at each time step for each surface patch using the Newton-Raphson method. Then, the patch surface temperature at the next time step ($T_i^{t+1}$) is solved using the energy balance equation (Eq. (1)), where the conduction term becomes:

$$Q_{G,i} = -k_1 A_i \frac{(T_i^{t+1} - T_{f,1}^t)}{0.5\Delta z_1} \tag{36}.$$

Here, $T_{f,1}$ (K), $k_1$ (W m$^{-1}$ K$^{-1}$), and $\Delta z_1$ (m), are the temperature, thermal conductivity, and thickness of the first fuel layer. From below, the solid or fuel is bounded by the soil and from the top, the surface of the patch is exposed to the atmosphere. In this way, in addition to the physical and material properties of the internal fuel layers, the model is regulated by the net radiation ($R_{net,i}$; Eq. (2)), local wind speed, ambient air temperature, precipitation, and relative humidity at the top surface (patch $i$), and by the soil moisture content and soil temperature at the bottom most fuel layer.

## 3. Full Model Validation

While the accurate formulation of each sub-model was verified and the performance of each was validated individually against analytical solutions or measured data, the holistic performance of TAMEFOE was tested against measurements from several experiments, data from different numerical solutions, and an analytical study reported in the literature. For the sake of brevity, only the full model tests and the moisture sub-model performance are presented here. In the first test, the model results for a simple transient 1D heat transfer problem are compared against the theoretical solution for a study reported by Yuan et al. (2020). For the second test, the model prediction was evaluated against field data (Voogt & Grimmond, 2000), as well as data from a 3D urban energy balance model, TUF-3D (Krayenhoff & Voogt, 2007), for a light industrial site in Vancouver, Canada under diurnally varying environmental conditions. In the third and fourth tests, the model prediction of fuel ignition time for fuels exposed to external heat sources was compared against laboratory experiments. Lastly, the performance of the model for predicting the fuel moisture content was evaluated against experiments by McCaw (1997).

**3.1 Validation against theoretical solution of transient 1D heat transfer**

The accuracy of the temperature evolution predicted by the model was first compared against the theoretical solution of a 1D heat transfer problem that was performed for a combustible solid as documented in Yuan et al. (2020). The problem considers a medium density fibreboard (MDF) with a square surface of dimensions 0.04 m × 0.04 m and a material thickness of 0.0184 m. The top surface was exposed to an external heat flux of 1 kW m$^{-2}$ and no heat loss was considered at the bottom most layer. The other relevant material and thermal properties of MDF used in the simulations are based on Yuan et al. (2020) and are listed in Table 1. No solar radiation was considered, the convection coefficient at the top surface was set to 10 W m$^{-2}$ K$^{-1}$, and the FDM was employed for the calculation of conduction. The initial and ambient air temperatures were set at 24.85°C, according to Yuan et al. (2020).

**Table 1** Thermal and material properties of medium density fibreboard (Yuan et al. 2020)

| Fuel Type | Density (kg m$^{-3}$) | Heat Capacity (J kg$^{-1}$K$^{-1}$) | Emissivity | Convection Coefficient (W m$^{-2}$K$^{-1}$) | Conductivity (W m$^{-1}$K$^{-1}$) | Material thickness (m) |
|---|---|---|---|---|---|---|
| Medium density fibreboard | 605 | 1340 | 0.86 | 10 | 0.15 | 0.0184 |

Figure 2 compares the temperature evolution of the top surface and the inner most layer of the sample between the model and the theoretical results. The results indicate that the temperature evolution at the surface exposed to external heat flux and the inner most material layer is well reproduced by TAMEFOE.

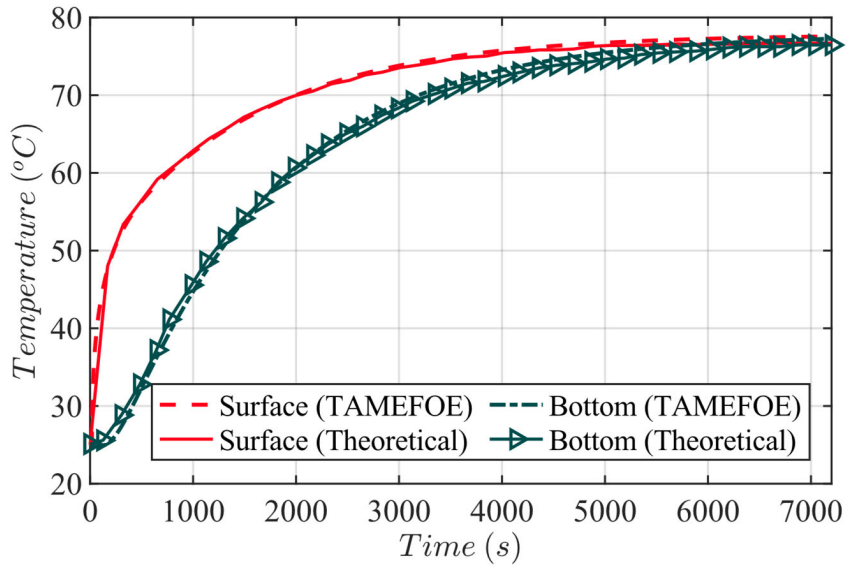

Figure 2: Comparison of the temperature evolution of the top surface and the inner most layer of the MDF sample between TAMEFOE and the theoretical results as documented in Yuan et al. (2020)

**3.2 Validation against diurnally measured surface temperature data from an industrial site**

The field experiment, reported in Voogt & Grimmond (2000) was performed in Vancouver, Canada in an industrial site consisting of one- to three-story buildings with a rectangular footprint, flat roofs, and no vegetation. The measurements were conducted on August 15, 1992, during the presence of well-defined sunlit and shaded patterns over the structures. Walls and street surface temperatures were measured using truck-mounted remote thermal sensors. While, for the roof, airborne thermal remotely sensed radiative surface temperatures were acquired. Further details about the site and measurements are available in Krayenhoff & Voogt (2007) and Voogt & Grimmond (2000). A 3-by-3 array of uniformly spaced buildings was configured for this validation study based on the geometrical ratios reported in Krayenhoff & Voogt (2007) and the results were analyzed for the central building unit as it was done in Krayenhoff & Voogt (2007). Besides field data, the model outputs were also compared against the outputs of the TUF-3D model. The building material's thermal properties (obtained from Krayenhoff & Voogt 2007) are collected in Table 2. In the absence of weather data (e.g., air temperature, wind speed, radiation) for the date of the measurements, the required weather condition information was extracted from the available TMY files. Using this weather data, the model was run over a period of four days from August 12 to August 15, with the first three days being considered as the spin-up period and the results from the last day were used for the comparisons. Here, the DOE-2 convection model was used for the vertical walls since it accounts for the wind directions local to each patch (see Appendix B for further details), the Simple Combined convection method was employed for the horizontal surfaces of the roofs and streets, and the Z-transform method was used for the conduction.

**Table 2** Wall, roof, and street material layers from the outer to the inner layer (based on Table 2 of Krayenhoff & Voogt, 2007)

| Material layers | Thickness (m) | Conductivity (W m$^{-1}$ K$^{-1}$) | Density (kg m$^{-3}$) | Specific heat (kJ kg$^{-1}$ K$^{-1}$) | Albedo | Emissivity |
|---|---|---|---|---|---|---|
| **Walls** ($T_{initial}$=19°C) | | | | | | |
| Layer 1 (outer) | 0.03 | 1.51 | 2400 | 0.88 | 0.5 | 0.9 |
| Layer 2 | 0.07 | 0.67 | 1600 | 0.625 | - | - |
| Layer 3 | 0.07 | 0.67 | 1600 | 0.625 | - | - |
| Layer 4 (inner) | 0.03 | 1.51 | 2400 | 0.88 | 0.5 | 0.9 |
| **Roof** ($T_{initial}$=12°C) | | | | | | |
| Layer 1 (outer) | 0.015 | 1.4 | 2000 | 0.88 | 0.12 | 0.92 |
| Layer 2 | 0.015 | 1.4 | 2000 | 0.88 | - | - |
| Layer 3 (Insulation) | 0.01 | 0.03 | 40 | 1.0 | - | - |
| Layer 4 (inner) | 0.03 | 1.51 | 2400 | 0.92 | - | - |
| **Street** ($T_{initial}$=20°C) | | | | | | |
| Layer 1 (outer) | 0.05 | 0.82 | 2110 | 0.82 | 0.08 | 0.95 |
| Layer 2 | 0.2 | 2.1 | 2400 | 0.83 | - | - |
| Layer 3 | 0.1 | 0.4 | 1300 | 1.08 | - | - |
| Layer 4 (inner) | 0.1 | 0.4 | 1300 | 1.08 | - | - |

Figure 3 compares the average surface temperatures of the north (N), south (S), east (E), and west (W) facing walls from TAMEFOE against the measured data and the TUF-3D modeled data. The results indicate that the diurnal changes in the wall surface temperatures throughout the day are well captured by TAMEFOE. This is further verified in Table 3, which shows the root mean square errors (RMSE) between the simulated and observed data. For comparison, the table also includes the RMSE for the TUF-3D model (obtained from Krayenhoff & Voogt 2007). The slight differences between the model output and observations are potentially related to the weather data input into the model, which was not provided in the original reference and, therefore, obtained from the TMY weather data files.

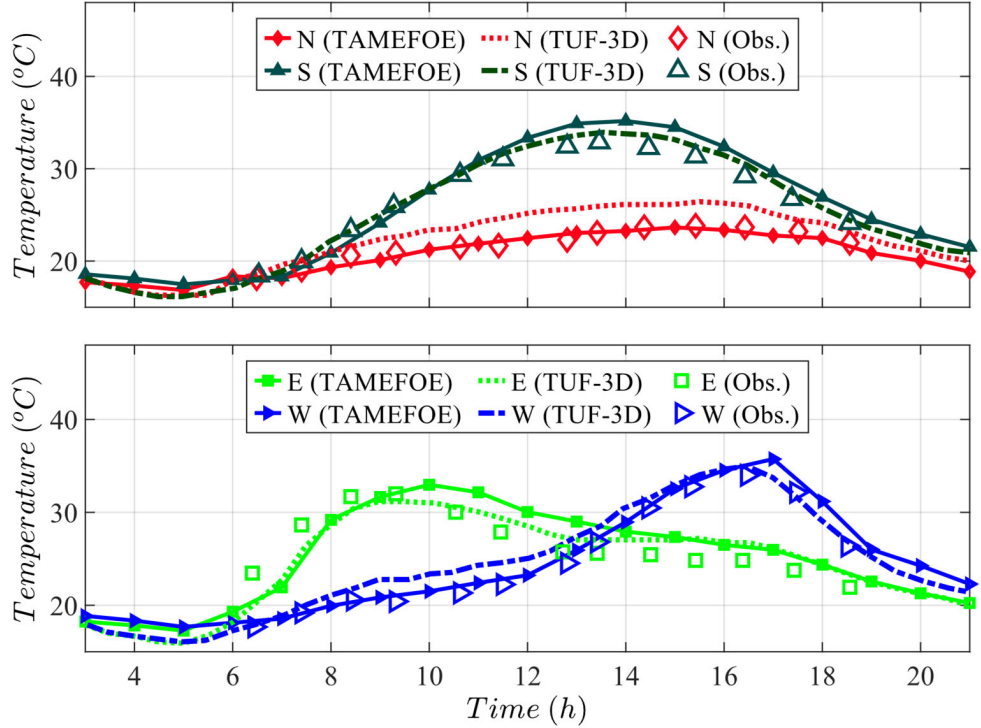

Figure 3: Comparison of TAMEFOE outputs against measured data (Obs.) and the simulated results of TUF-3D for facet-average surface temperatures of the north (N), south (S), east (E), and west (W) facing walls throughout the day for Aug 15, 1992, in the Vancouver Light Industrial site.

Table 3 Root Mean Square Errors (RMSE) between simulated surface temperatures and observations for the Vancouver Light Industrial site. RMSE of the TUF-3D model is shown for comparison.

| Material layers | N - wall | S - wall | E - wall | W - wall |
|---|---|---|---|---|
| TAMEFOE | 0.5 | 1.5 | 2.2 | 0.9 |
| TUF-3D | 2.2 | 0.8 | 1.9 | 1.6 |

Figure 4 compares the diurnal changes of the average roof and street surface temperatures between the observed data and TAMEFOE and TUF-3D modeled data. It can be seen that the simulated results of TAMEFOE for the roof surface temperatures compare well with the field observations and show a better prediction compared to that of TUF-3D. A maximum difference of 3.28°C compared to 12.97°C of the TUF-3D model is seen between the field observations and TAMEFOE. The underperformance of TUF-3D, in this case, was related to the uncertainty in the surface roughness length (that is used in the Monin–Obukhov similarly theory (MOST) formulation of the convection model used in TUF-3D for horizontal surfaces), insulation layer thickness, and radiative parameters (Krayenhoff & Voogt, 2007). In the current model, the same input for the insulation layer thickness is used. However, the Monin–Obukhov similarly theory (and thus the surface roughness length) is not employed in TAMEFOE for the convection modeling, and the radiative parameters and the air temperature information are obtained from TMY files rather than from helicopter reading (due to the absence of the information). The simulated street temperatures compare well with the observations in the morning; however, they are underestimated after midday with a maximum difference of 9.86°C compared to 5.34°C of the TUF-3D model. This difference is also potentially related to the use of different sources for the local weather data in the two models.

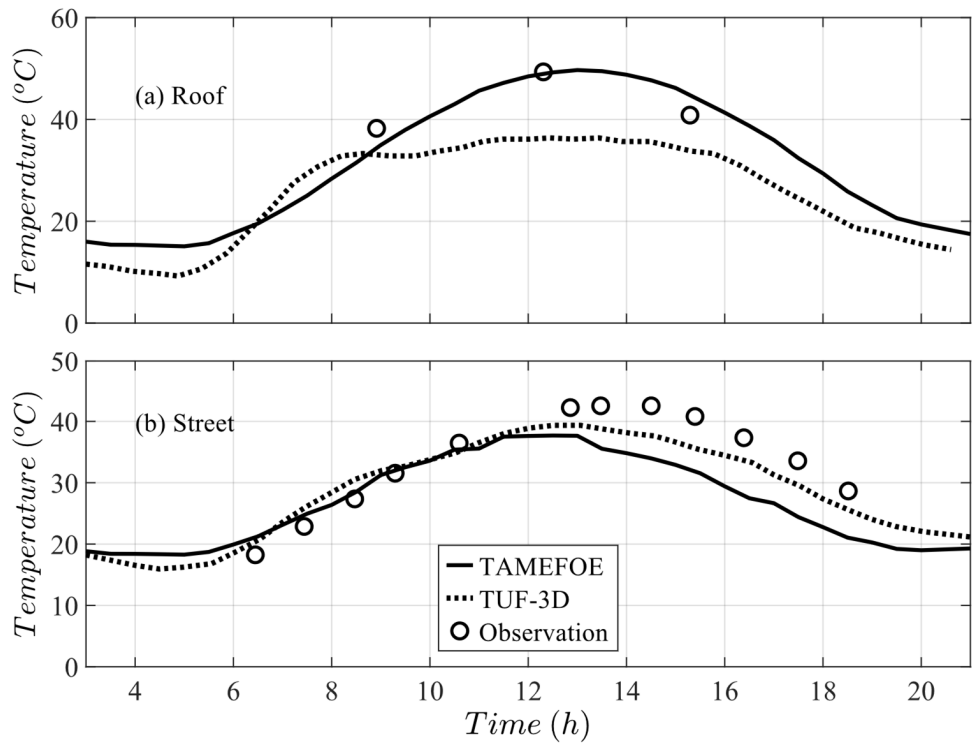

Figure 4: Comparison of TAMEFOE outputs against measured data (Obs.) and simulation results of TUF-3D for facet-average surface temperatures of the (a) roof and (b) street throughout the day of Aug 15, 1992, in the Vancouver Light Industrial site.

### 3.3 Validation against experimentally measured fuel ignition time

To test the accuracy of the model for predicting the temperature evolution of objects subjected to high external heat, the model results were compared against the measured data of two piloted ignition experiments of wildland fuel beds (Mindykowski et al. 2011; Rivera et al. 2021). The fuel beds in both of these controlled experiments were made of combustible materials that were heated by either smoldering firebrands (Rivera et al. 2021) or infrared heaters (Mindykowski et al. 2011). In the experimental setup by Mindykowski et al. (2011), dead and dried Maritime Pine needles were used and placed in a cylindrical steel basket holder with a diameter of 0.126 m and a height of 0.03 m. Longer needles were cut using scissors and all the needles were randomly placed by hand within the basket to reach a desired volume fraction ($V_{frac}$) in four different experiments with different incident heat fluxes. The heat was generated using infrared heaters that were located at a distance of 0.073 m above the fuel basket. The imposed heat ranged from 12.5 kW m$^{-2}$ to 30 kW m$^{-2}$ in increments of 2.5 kW m$^{-2}$ and the time to ignition was recorded at the visual appearance of flaming combustion for each incident heat flux. The time of flaming was also verified using a recorded video. More information on the experimental setup can be found in Mindykowski et al. (2011). To numerically emulate this experiment, a similar setup was considered in the computational model. However, since information on the random setup of the fuel elements was not provided, a solid material with equivalent effective density was considered in TAMEFOE. The fuel conductivity was set to the conductivity of the pinewood material (0.12 W m$^{-1}$ K$^{-1}$; Baranovskiy & Malinin 2020) as that of the Maritime Pine needles was not provided by Mindykowski et al. (2011) or found elsewhere. The other relevant material and thermal properties of the Maritime Pine needle fuel bed used in the simulations are based on Mindykowski et al. (2011) and listed in Table 4. The experiments were done indoors, therefore no solar radiation was considered in the numerical model. For the convection model, due to the absence of relevant parameter information for calculating the convection coefficient, this coefficient was set to 15 W m$^{-2}$ K$^{-1}$ (based on the theoretical solution presented in Rivera et al. (2021)) and the ambient air temperature was set to a typical room temperature of 25°C.

Table 4 Thermal and material properties of wildland fuel beds for Maritime Pine needles (Mindykowski et al. 2011) and Monterey Pine needles (Rivera et al. 2021)

| Fuel Type | Volume fraction | Density (kg m$^{-3}$) | Heat Capacity (J kg$^{-1}$K$^{-1}$) | Ignition Temp. (K) | Emissivity | Convection Coefficient (W m$^{-2}$K$^{-1}$) | Equivalent Density (kg m$^{-3}$) | Conductivity (W m$^{-1}$K$^{-1}$) | Total material thickness (m) |
|---|---|---|---|---|---|---|---|---|---|
| Maritime Pine needles | 0.08 | 630 | 1470 | 400 | 0.8646 | 15 | 51.53 | 0.12 | 0.036 |

| | | | | | | | | |
|---|---|---|---|---|---|---|---|---|
| Monterey Pine needles | 0.09 | 615.3 | 1470 | 549 | 0.8646 | 15 | 56.49 | 0.12 | 0.045 |

Figure 5 compares the time taken for the average temperature of the entire fuel layer in the sample holder to reach the ignition temperature ($t_{ignition}$) between the simulation and experiment for the provided range of incident heat fluxes. The figure shows a good agreement between the model and experimental data for high incident heat fluxes. However, ignition times for the lower range of heat fluxes are underestimated. This underestimation is potentially related to the comparison of the average temperatures of the entire material layer (as it was done in the experiment), which is not completely accurate for the lower incident heat fluxes since the needles closer to the heat source are likely to reach the ignition temperatures sooner. Other sources of differences could be related to the use of the fuel conductivity for the pinewood (rather than Maritime Pine needle) material or due to the differences in the experimental and numerical setups. In the experiments, the pine needles were distributed randomly, while the numerical setup employs equivalent density of the combined setup that includes the needles and the air gaps. Equivalent properties were used in these simulations since the dimensions of each needle in the fuel layer and their random setup were not provided, which makes it impractical to accurately replicate the exact experimental setup.

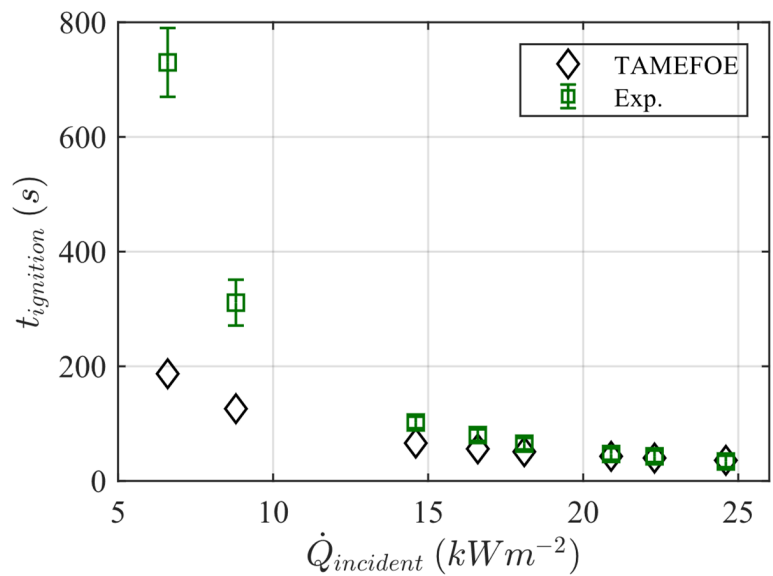

Figure 5: Comparison of the ignition time as a function of incident heat flux between the TAMEFOE results and the experimental data from Mindykowski et al. (2011) for a Maritime Pine needle fuel layer

The second validation study for objects exposed to high external heat was conducted using the experimental data of Rivera et al. (2021) in which a set of ignition experiments were carried out on an idealized firebrand-ignition test bench-scale apparatus with dead and dried Monterey Pine needles. The experimental setup represented a relatively complex-shaped setting consisting of a cylindrical-shaped sample holder with a radius of 0.065 m and a height of 0.046 m, with an axisymmetric void at the center with a radius of 0.02m. A small cylindrical-shaped heater element (representing an idealized firebrand) was placed vertically at the center of the void to provide varying incident heat flux to the fuel layer. Similar to the experiments by Mindykowski et al. (2011), the pine needles were randomly placed inside the sample holder by hand to achieve a desired volume fraction for the fuel layer. The time to ignition of the fuel layer in this setup was recorded at the initial appearance of a flame. A similar setup was used in TAMEFOE to evaluate the numerical model performance. As in the experiments, the inner surface of the hollow cylinder was exposed to various heat fluxes ranging from 6.6 kW m$^{-2}$ to 24.6 kW m$^{-2}$. The equivalent density and thermal and material properties of the fuel layer are presented in Table 4 based on Rivera et al. (2021). Similar to the validation study against the Mindykowski et al. (2011) case, the convection coefficient was fixed at 15 W m$^{-2}$ K$^{-1}$ (Rivera et al. 2021), the external air temperature was set at 25°C, and the FDM was employed for the calculation of conduction.

Figure 6 compares the numerical results of TAMEFOE against measured data of $t_{ignition}$ (i.e., the time taken for the average temperature of the fuel layer to reach the ignition temperature) for various incident heat fluxes. The results show a good performance of the model, except for the lowest incident heat flux, where the highest experimental error was reported as well. As also stated by Rivera et al. (2021), the experiments at lower heat fluxes tend to have higher errors as the needles are likely to be consumed before ignition is attained. The lower ignition time prediction by the model is likely to be related to the comparison of the average material layer temperature against that of the experiment. Analogous to the previous validation study, the other minor differences between the model and the experiment are likely to be related to the use of equivalent material density since the exact random setup of the fuel needles within the sample holder is not achievable.

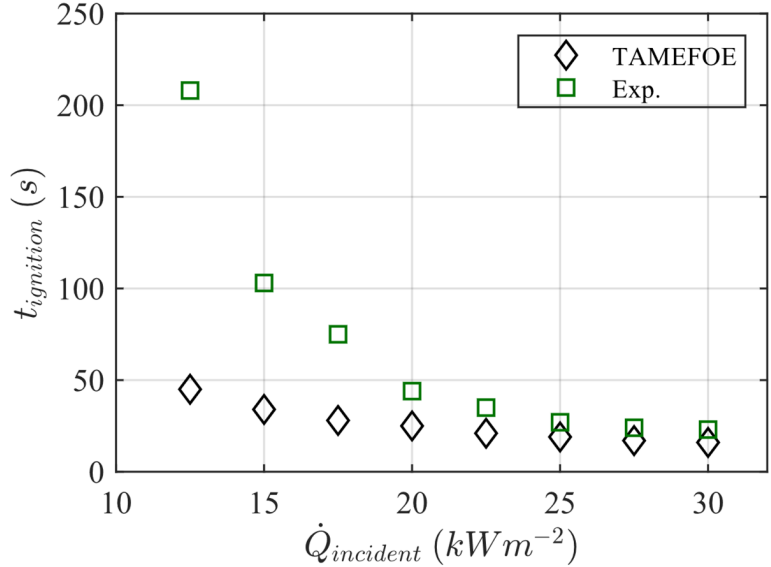

Figure 6: Comparison of ignition time as a function of incident heat flux between the model and experiments from Rivera et al. (2021) on the Monterey Pine needle fuel layer.

### 3.4 Validation of the fuel moisture model

The accuracy of the fuel moisture model was examined in comparison to fuel moisture measurements made in a Eucalyptus mallee scrub near Perup in southwest Western Australia, reported in McCaw (1997). The scrubs were 4 m tall with 40% canopy closure and the fuel layer under consideration was 0.01 m – 0.02 m deep. The fuel moisture samples and the corresponding weather conditions were collected over 4 days from 15 to 18 April 1996 by the Western Australian Department of Conservation and Land Management (McCaw, 1997). The fuel moisture content (FMC) was determined by oven drying. Numerically, similar to Matthews (2006), five fuel layers of equal depth (0.004 m each) were considered with the same material and thermal properties (Appendix D). Except for temperature and relative humidity (Fig. 4a of Matthews, 2006), the remaining required weather data inputs (i.e., pressure, solar radiation, and wind speed) that were not provided by Matthews (2006) were obtained from the TMY data file for Perup. The soil temperature was considered to be equal to the air temperature and the water vapor flux from the soil was set to zero, according to Matthews (2006).

Figure 7 compares the FMC between TAMEFOE and the experiments of McCaw (1997) and the model results from Matthews (2006). In general, the results show a good performance of the model. When compared with the experiments, the simulation results from TAMEFOE reveal differences, especially for the FMC on 17 April, while the model results from Matthews (2006) provide a better prediction of the minimum and maximum FMC for 16 April and 17 April. These disparities in the TAMEFOE predictions are due to the input weather parameters that were not available for the day of the experiment and had to be obtained from the TMY file.

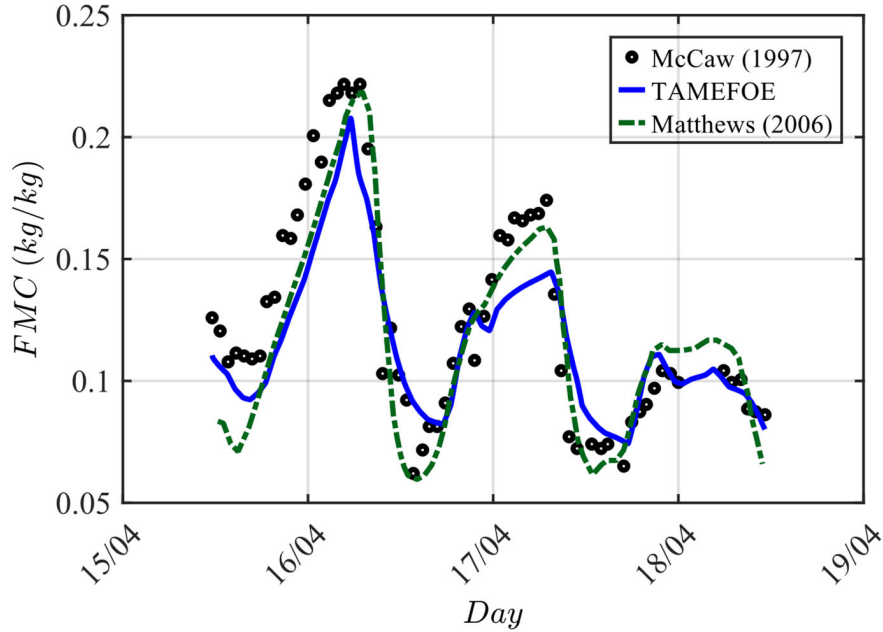

Figure 7: Comparison of the fuel moisture content (FMC) between TAMEFOE and the experiments from McCaw (1997) and model results from Matthews (2006).

## 4. Model Application: Effects of Surface Topography and Material on Diurnal Fuel Ignition Potential

The fuel ignition potential to fire and ember is determined by the fuel temperature and dryness, which are spatiotemporally variable depending on the fuel material, timing of the day/year, local weather conditions, terrain and topography, geographical location, and vicinity to fire flame (if any exists). In this paper, which focuses on the model description, we aim to introduce a simple example of the application of the model. In this example, we investigate how topography and material influence the fuel ignition potential throughout a day in a small domain. We choose the weather condition of November 8, 2018, and geographical location of Paradise, CA (the day and location of the Camp Fire) for this example study. More detailed investigations of the application of TAMEFOE to predict fuel ignition potential are left for future studies.

### 4.1 Simulation setup

With hills, mountain ridges, vegetations, residential buildings made of different materials, and roads, the landscape encompassing Paradise, CA, is intricate. To address a complex fuel layout, in this brief example, we chose a small domain with a complex hilly structure and a sparse arrangement of a few buildings with different heights. While TAMEFOE accounts for multiple materials within the same domain, for simplicity, this example only uses two material types to represent the terrain. The buildings and pavement are made of concrete while the landscape is created using vegetative fuel, as shown in Fig. 8. The thermal and physical properties of these materials are listed in Table 5.

The local weather data was taken from the TMY weather data file for Paradise, CA location (39.77° N, 121.62° W, and 467 m altitude). The simulations in TAMEFOE were performed for a period of 4 days (5 – 8 November 2018). The first three days were considered as the model spin-up period and the results from the last day, November 8, were analyzed. Based on the weather data file, the minimum and maximum air temperatures for this day were 7.7 °C and 20.1 °C, respectively, with an average wind speed of 2.6 m s$^{-1}$, and an average relative humidity of 17.4 %. The convection coefficient was set at 10 W m$^{-2}$ K$^{-1}$ and the soil temperature was calculated based on the deep soil temperature (Eq. 29; Hillel (1982)).

**Table 5** Thermal and material properties of wildland fuel beds (based on Table 1 of Matthews, 2006) and concrete (based on Table 2 of Krayenhoff & Voogt, 2007)

| Fuel Type | Density (kg m$^{-3}$) | Heat Capacity (J kg$^{-1}$K$^{-1}$) | Emissivity | Bulk Density (kg m$^{-3}$) | Conductivity (W m$^{-1}$K$^{-1}$) | Total material thickness (m) |
|---|---|---|---|---|---|---|
| Vegetation | 550 | 1004.5 | 0.9 | 62 | 0.14 | 0.05 |
| Concrete | 2400 | 920 | 0.9 | 2100 | 1.51 | 0.03 |

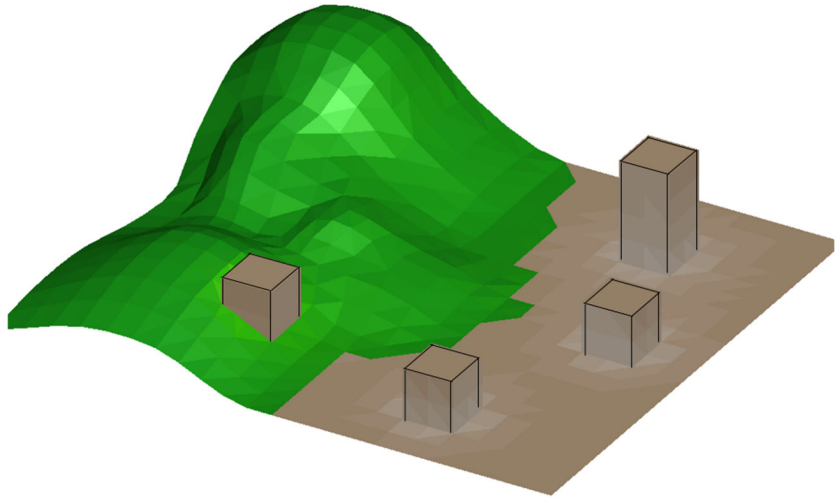

Figure 8: Schematic of the simulation setup. The brown and green regions, respectively, indicate areas with concrete and vegetative materials.

**4.2 Results and discussions**

Figure 9 shows snapshots of the surface temperature and moisture content contours at two times of the day, morning, at 0800 LT (local time = UTC − 8 h) and afternoon, at 1300 LT. The figure illustrates the dynamically changing sunlit and shaded regions over the terrain (through the temperature contours), as well as the corresponding changes in the moisture levels. The minimum daily moisture content level on November 8, 2018, in this terrain is in the range of 0.0039−0.0105 kg kg$^{-1}$, happening at 1130 LT. This minimum moisture level happens at the top of the largest hill, where the temperature has the highest values (47.60−55.79°C). From the simulated spatiotemporal variations in the moisture content levels and surface temperatures, it is possible to identify the areas within the domain that are more susceptible to fire (have high ignition potentials). These are the vegetated regions that have the least moisture content, are not shaded, and receive the most radiation during the time in question.

According to the California Department of Forestry and Fire Protection's report on the Green Sheet Camp Incident (CAL Fire, 18-CA-BTU-016737), the fuel moisture levels were observed to be at 5% (kg kg$^{-1}$) at a location southeast of the Camp fire, while the average fuel moisture level of the region for the month of November was reported at 17% in the same document. TAMEFOE's simulated moisture level for November 8, 2018, at the geographical location of Paradise, CA, is within the same range as the measured values, as shown in Figs. 9b and d. These moisture levels indicate how prone the fuels can be to fire and embers. Burning embers that fall on such areas have the potential to start spot fires. Indeed, as mentioned earlier, in Camp fire, numerous spot fires were ignited prior to the arrival of the main fire front due to the receptivity of the fuels (Maranghides et al., 2021).

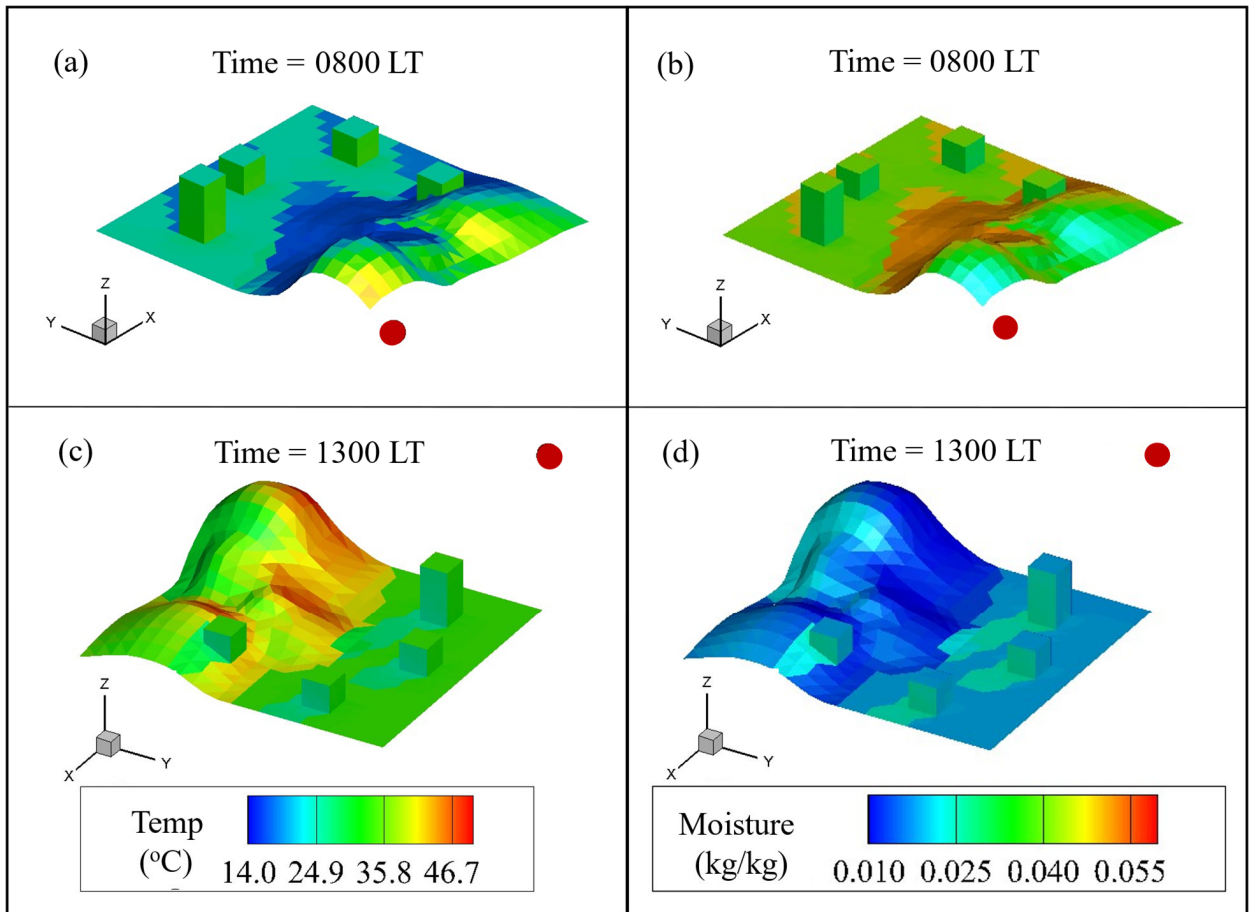

Figure 9: Snapshots of the simulated surface temperature (a, c) and moisture content (b, d) at 0800 LT and 1300 LT, respectively, for the geographical location of Paradise, CA on November 8, 2018. For a better view, the snapshots are provided from two angles. The red circle in the figures indicates the instantaneous location of the sun with respect to the terrain at the corresponding hours.

## 5. Model Sensitivity to Parameters

The sensitivity of the model to its parameters was investigated by computing the root mean square deviations in surface temperature and fuel moisture content outputs from the corresponding base case results. The moisture model validation setup, which corresponds to a four-day scenario, served as the basis for this analysis. This examination was based on an arbitrarily chosen perturbation of $\pm 20\%$ to each of the model parameters. The model parameters that were perturbed were wind speed, direct normal irradiation, fuel conductivity, ambient air temperature, relative humidity, surface albedo, specific heat capacity of fuel, density of fuel, latitude, deep soil temperature, Nelson (1984) model parameters A and B, fuel surface conductance, fuel surface area to volume ratio, soil humidity, and bulk density. Table 6 displays the model sensitivity results to each parameter in terms of average values of the model sensitivity to $\pm 20\%$ change of each parameter.

As expected, surface temperature is very sensitive to changes in ambient air temperature, solar irradiation, latitude, and wind speed, which are all based on the geographical location of the fuel and the time of the year. The results indicate that surface temperature has a weaker sensitivity to surface albedo, deep soil temperature, and conductivity (in that order). Sensitivity to all other parameters (i.e., bulk density, specific heat capacity, Nelson (1984) model parameter A and B, fuel surface conductance, fuel surface area to volume ratio, density, specific humidity of soil, ambient relative humidity) is small.

As also indicated by Matthews (2006), the Nelson (1984) model parameters A and B have the greatest influence on the fuel moisture content. The third most critical factor impacting the fuel moisture content is the ambient relative humidity. As can be estimated, an increase (decrease) in the ambient relative humidity leads to a considerable increase (decrease) in the moisture levels. The fuel moisture content exhibits a weaker sensitivity to the deep soil temperature, solar irradiance, latitude, ambient air temperature, soil's specific humidity, wind speed, and conductivity (in that order). Table 6 demonstrates that there is a little sensitivity to all other parameters, including albedo, bulk density, soil specific humidity, density, fuel surface area to volume ratio, and fuel surface conductance.

**Table 6** Sensitivity of the fuel temperature ($T_f$) and fuel moisture content ($m_f$) to various model parameters. The units of sensitivity are °C (temperature) and kg kg$^{-1}$ (moisture). The reference values are based on the moisture model validation case. Mean values of the base case fuel temperature and moisture content over the course of four days are also provided in the table as a reference.

| Parameter | $m_f$ (kg kg$^{-1}$) | $T_f$ (°C) |
|---|---|---|
| Wind speed (m s$^{-1}$) | 0.00114 | 0.71733 |
| Direct normal irradiation (W m$^{-2}$) | 0.00227 | 1.26873 |
| Conductivity (W m$^{-1}$ K$^{-1}$) | 0.00111 | 0.12779 |
| Ambient air temperature (°C) | 0.00135 | 3.11787 |
| Ambient relative humidity (%) | 0.01738 | 0.00060 |
| Albedo | 0.00057 | 0.31718 |
| Specific heat capacity (J kg$^{-1}$ K$^{-1}$) | 0.00031 | 0.01734 |
| Density (kg m$^{-3}$) | 0.00002 | 0.00107 |
| Latitude | 0.00144 | 0.78557 |
| Deep soil temperature (°C) | 0.00454 | 0.25369 |
| Nelson (1984) model parameter A | 0.05474 | 0.00108 |
| Nelson (1984) model parameter B | 0.02330 | 0.00029 |
| Fuel surface conductance (m s$^{-1}$) | 0.00002 | 0.00108 |
| Fuel surface area to volume ratio (m$^{-1}$) | 0.00002 | 0.00108 |
| Specific humidity of soil (kg kg$^{-1}$) | 0.00120 | 0.00104 |
| Bulk density (kg m$^{-3}$) | 0.00032 | 0.01810 |
| **Mean values in the base case** | **0.10740** | **19.57419** |

# 6. Conclusions

A new high-resolution coupled surface energy and moisture balance model, Temperature And Moisture Evolution predictor for complex Fuel in Open Environment (TAMEFOE), is developed to predict the spatiotemporal surface and body temperature and moisture content evolution of 3D objects and fuels in diurnally variable environmental conditions. One of the motivations for developing TAMEFOE was the lack of computational capabilities to predict vulnerability of combination of complex shaped/material fuels to ignition in different weather, solar, and environmental conditions. Such a capability is particularly important in fire-safety and wildfire-propagation research in complex environments (e.g., wildland-urban interfaces) and for determining when complex multi-material objects/fuels with particular properties reach the ignition point if they are exposed to fire heat at different times of the day.

The model uses nonuniform triangular surface patches to handle complex 3D geometries. The objects under investigation can be of any random shape and made of multiple materials and material layers. The discrete, high spatiotemporal resolution surface temperature and moisture content information of TAMEFOE can also be used to provide unsteady boundary conditions for computational fluid dynamics simulations when coupled physics is under investigation.

In TAMEFOE, the environmental attributes and thermal and material properties of each sub-surface (or patch) are local to each patch, and the energy and moisture balances are solved at the patch scale, using a thorough treatment of radiation exchanges at the same scale. A ray-tracing shadow sub-model has been developed and incorporated into the model to represent the spatiotemporal variations of the shadow distributions in high resolution in case any exists. The open-environment weather and solar radiation data in the model are based on typical meteorological year (TMY) weather data files, which are available for each geographical location.

Following tests of the sub-models, the performance of the whole model for diurnally variable environmental conditions and surface temperatures has been validated against field measurements (Voogt & Grimmond 2000) and modeled surface temperatures from an urban energy balance model (Krayenhoff & Voogt 2007), displaying close agreements with both. The model performance has also been evaluated for objects under the influence of high external heat sources using the results of two controlled experiments (Mindykowski et al. 2011; Rivera et al. 2020). The performance of the model has also been tested for a simple transient 1D heat transfer problem against an analytical solution. Lastly, the fuel moisture experiments of McCaw (1997) were used to assess the model's performance for the diurnal evolution of fuel moisture content. The results of the validation studies indicate the accurate performance of the model in predicting the temperature and moisture content evolution of objects and fuels under different conditions. Due to the lack of other measured data with necessary details, the model tests were limited to the cases presented in Sect. 3. However, more tests are desired to indicate potential model weaknesses and the range of model performance under different environmental, material, and geometrical conditions. The model can be improved in several aspects. For example, the conduction model is currently limited to one dimension, which makes the model not readily applicable to very small objects. The computational costs of the simulations depend on the model geometry resolution (number of patches). For a very high-resolution large domain, as expected, the computational cost can be noticeable, and therefore, improvement of the model to run in parallel across distributed memory using a message passing interface is desirable.

## Acknowledgment

This work is supported by funding from the Department of Defense Strategic Environmental Research and Development Program (SERDP) RC20-1298 and Florida State University. We also thank Iago Dal-Ri Dos Santos for his help with the surface topography modelling in the model application section.

# Appendix A View Factor Model

The view factor model used in this study employs the contour integral (Eq. (38); Sparrow 1963) representation of the well-known form (Eq. (38); Incropera et al. 2007) of the view factor ($F_{QP}$) between two patches $Q$ and $P$:

$$F_{QP} = \frac{1}{A_Q} \int \int \frac{\cos\theta_Q \cos\theta_P}{\pi R^2} dA_Q \, dA_P \tag{37},$$

$$F_{QP} = \frac{1}{2\pi A_Q} \oint \oint \ln S \, \mathbf{ds_Q} \, \mathbf{ds_P} \tag{38},$$

where $\theta_Q$ is the angle between the outward normal of the patch $Q$ and the line joining patches $P$ and $Q$, $\theta_P$ is the angle between the outward normal of the patch $P$ and the line joining patches $P$ and $Q$, $R$ (m) is the distance between the two patches. $A_Q$ (m²) and $A_P$ (m²) are the areas of patches $P$ and $Q$, respectively, and $S$ is the distance between two differential line vectors ($\mathbf{ds_Q}$ and $\mathbf{ds_Q}$) as shown in Fig. 10. This contour formula is then transformed into a vector parametric representation (Mazumder & Ravishankar 2012):

$$\begin{aligned} F_{QP} = \frac{1}{4\pi A_Q} \sum_{n=1}^{N} \sum_{m=1}^{M} \int_0^1 \int_0^1 &\ln\left(\lambda_Q^2 |\mathbf{q}_{m,m+1}|^2 + \lambda_P^2 |\mathbf{p}_{n,n+1}|^2 + |\mathbf{Q}_m \mathbf{P}_n|^2 \right. \\ &- 2\lambda_Q \, \mathbf{Q}_m\mathbf{P}_n \cdot \mathbf{q}_{m,m+1} + 2\lambda_P \, \mathbf{Q}_m\mathbf{P}_n \cdot \mathbf{p}_{n,n+1} \\ &\left. - 2\lambda_Q \lambda_P \, \mathbf{p}_{n,n+1} \cdot \mathbf{q}_{m,m+1} \right) \mathbf{p}_{n,n+1} \cdot \mathbf{q}_{m,m+1} \, d\lambda_Q \, d\lambda_P \end{aligned} \tag{39}.$$

In this equation, $\mathbf{p}_{n,n+1}$ and $\mathbf{q}_{m,m+1}$ are vectors on the contours of polygons $P$ and $Q$, respectively, directed from one vertex of the polygon to the other. $\mathbf{Q}_m\mathbf{P}_n$ is a vector from a vertex on polygon $Q$ to a vertex on polygon $P$. $0 < \lambda_P < 1$ and $0 < \lambda_Q < 1$ represent a fraction of the vectors $\mathbf{p}_{n,n+1}$ and $\mathbf{q}_{m,m+1}$. The integral in Eq. (40) is calculated using the 10-point Gauss-Legendre quadrature scheme (Appendix B) (Beyer 1996). If an edge is shared between the polygons, the following relation is used

$$F_{QP\_Shared} = |\mathbf{q}_{shared\,edge}|^2 \left(1.5 - 0.5\ln\left(|\mathbf{q}_{shared\,edge}|^2\right)\right) \tag{40},$$

Where, $|\mathbf{q}_{shared\,edge}|$ is the length of the shared edge. Please refer to Mazumder and Ravishankar (2012) for further details.

The view factor implementation in this model was verified with the cases presented in Mazumder & Ravishankar (2012) and Hoff & Janni (1989) with excellent accuracy.

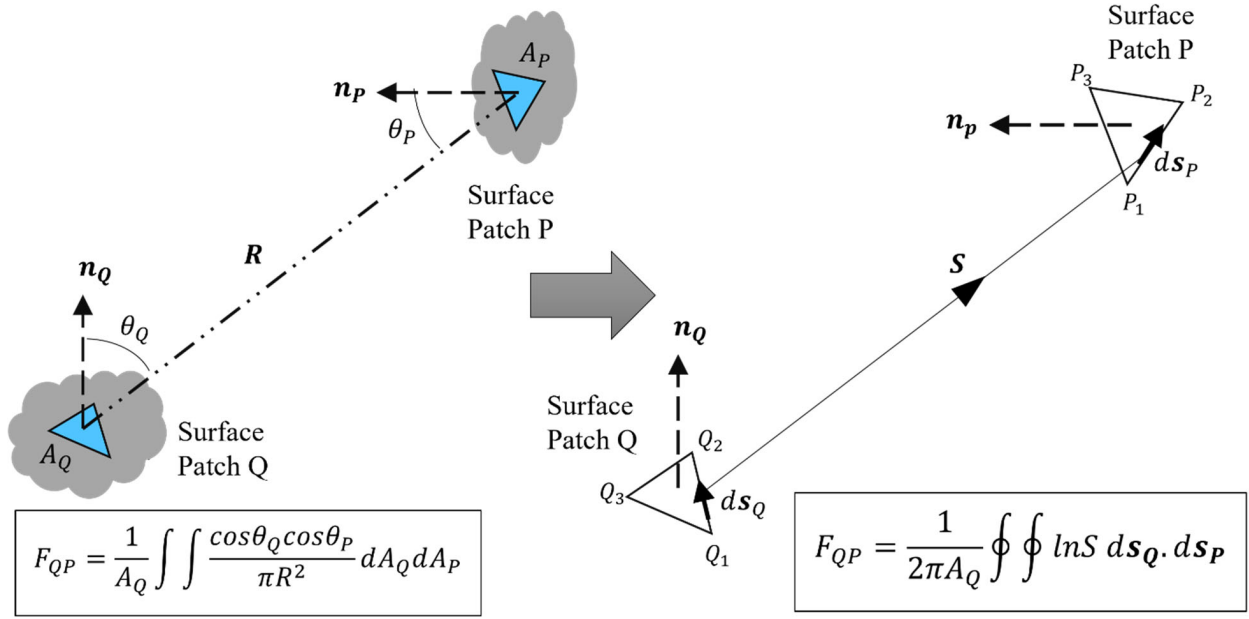

Figure 10: Diagrammatic representation of the conversion of the well-known view factor form to a vector parametric contour integral form for a triangular patch.

## Appendix B Gauss-Legendre Quadrature Scheme

To evaluate the integrals in Eq. (40), the following scheme is used (Beyer 1996):

$$\int_a^b f(x)dx = \sum_{i=1}^{\infty} w_i f(x_i) \approx \sum_{i=1}^{n} w_i f(x_i) \qquad (41).$$

Here, $f(x)$ is the integral to be evaluated, $(a, b)$ are the limits of the integration, $n$ is the number of points used in the scheme, $w_i$ are the weights or coefficients, and $x_i$ are the nodes or abscissa. The weights ($w_i$) and nodes ($x_i$) are chosen to achieve maximum accuracy. Here, a 10-point Gaussian quadrature scheme is used ($n = 10$). The weights and nodes used in this scheme are shown in Table 7. This scheme requires the integral limits to be within the range (-1,1), which entails the transformation of the coordinates in the following manner:

$$\int_a^b f(x)dx = \frac{(b-a)}{2}\int_{-1}^{1} f\left(\frac{b-a}{2}x_i + \frac{b+a}{2}\right)dx$$
$$\approx \frac{(b-a)}{2}\sum_{i=1}^{n} w_i f\left(\frac{b-a}{2}x_i + \frac{b+a}{2}\right) \qquad (42).$$

**Table 7** The weights ($w_i$) and nodes ($x_i$) used for a 10-point Gaussian quadrature scheme

| $i$ | weight - $w_i$ | abscissa - $x_i$ |
|---|---|---|
| 1 | 0.295524225 | -0.148874339 |
| 2 | 0.295524225 | 0.148874339 |
| 3 | 0.269266719 | -0.433395394 |
| 4 | 0.269266719 | 0.433395394 |
| 5 | 0.219086363 | -0.679409568 |
| 6 | 0.219086363 | 0.679409568 |
| 7 | 0.149451349 | -0.865063367 |
| 8 | 0.149451349 | 0.865063367 |
| 9 | 0.066671344 | -0.973906529 |
| 10 | 0.066671344 | 0.973906529 |

# Appendix C Convection Methods

## Nusselt Jurgess

This method constitutes of a large constant that represents the natural convection part of the total convection coefficient. The method can be applied to surfaces/patches with any orientation or roughness. This method takes the following simplified form in the SI units (Nusselt & Jurges 1922; Palyvos 2008):

$$h = 5.8 + 3.94 V_z \qquad (43),$$

where $h$ (W m$^{-2}$ K$^{-1}$) is the convection coefficient and $V_z$ (m s$^{-1}$) is the local wind velocity, which is adjusted according to the height of the surface patch's centroid.

## McAdams

This method has similar characteristics to the Nusselt Jurgess method. The form of this equation in the SI units is (McAdams 1954; Palyvos 2008):

$$h = 5.7 + 3.8 V_z \qquad (44).$$

## The Simple Combined Method

This method uses the local wind velocity ($V_z$ (m s$^{-1}$)) and the surface roughness parameters ($D$ (W m$^{-2}$ K$^{-1}$), $E$ (J m$^{-3}$ K$^{-1}$), and $F$ (J s m$^{-4}$ K$^{-1}$)) to calculate the convection coefficient (EnergyPlus 2021):

$$h = D + E V_z + F V_z^2 \qquad (45).$$

Table 8 Roughness coefficients for the Simple Combined Method (ASHRAE 1989)

| Roughness Index | D (W m⁻² K⁻¹) | E (J m⁻³ K⁻¹) | F (J s m⁻⁴ K⁻¹) |
|---|---|---|---|
| 1. Very Rough | 11.58 | 5.894 | 0.0 |
| 2. Rough | 12.49 | 4.065 | 0.028 |
| 3. Medium Rough | 10.79 | 4.192 | 0.0 |
| 4. Medium Smooth | 8.23 | 4.0 | -0.057 |
| 5. Smooth | 10.22 | 3.1 | 0.0 |
| 6. Very Smooth | 8.23 | 3.33 | -0.036 |

**TARP**

Thermal Analysis Research Program (TARP) model is a comprehensive convection model, which combines a correlation from ASHRAE and the flat plate experiments by Sparrow et al. 1979 (Walton 1983). In this model, the convection is split into forced and natural components, where the forced component (Eq. (47)) is based on a correlation by Sparrow et al. 1979:

$$h = h_{natural} + h_{forced} \tag{46},$$

$$h_{forced} = 2.537 W_f R_f \left(\frac{PV_z}{A}\right)^{0.5} \tag{47}.$$

Here, $h_{natural}$ (W m⁻² K⁻¹) is the natural convection coefficient, $h_{forced}$ (W m⁻² K⁻¹) is the forced convection coefficient, and $V_z$ (m s⁻¹) is the local wind velocity in m s⁻¹, which is adjusted according to the height of the surface patch's centroid. $A$ (m²) is the area of the surface, $W_f$ is equal to 1 for the windward surfaces and 0.5 for the leeward surfaces, $R_f$ is a surface roughness multiplier, and $P$ (m) is the perimeter of the surface.

Table 9 Surface roughness multipliers (Walton 1981)

| Roughness Index | $R_f$ |
|---|---|
| 1. Very Rough | 2.17 |
| 2. Rough | 1.67 |
| 3. Medium Rough | 1.52 |
| 4. Medium Smooth | 1.13 |
| 5. Smooth | 1.11 |
| 6. Very Smooth | 1.0 |

The natural component is calculated based on the following: For $\Delta T = 0$ or vertical surfaces, where $\Delta T$ is the temperature difference between the surface and air:

$$h_{natural} = 1.31 \, |\Delta T|^{1/3} \tag{48}$$

For $\Delta T > 0$ and a downward-facing surface or $\Delta T < 0$ and an upward-facing surface:

$$h_{natural} = \frac{9.482|\Delta T|^{1/3}}{7.283 - |\cos \epsilon|} \tag{49}$$

For $\Delta T > 0$ and an upward-facing surface or $\Delta T < 0$ and a downward-facing surface:

$$h_{natural} = \frac{1.81|\Delta T|^{1/3}}{1.382 + |\cos \epsilon|} \tag{50}$$

Here, $\epsilon$ is the surface tilt angle between the ground outward normal and the surface outward normal.

**The DOE-2 Method**

The DOE-2 convection model is the combination of the MoWiTT and BLAST Detailed models (LBL 1994). The BLAST model is similar to the TARP model with the calculation of the local wind velocity ($V_z$) being the only difference between the two (Mirsadeghi et al. 2013). The MoWiTT algorithm applies to smooth, vertical surfaces and has the following form:

$$h_c = \sqrt{(C_1(\Delta T)^{1/3})^2 + (aV_z^b)^2} \tag{51}$$

Here $h_c$ (W m$^{-2}$ K$^{-1}$) is the convection coefficient for very smooth surfaces, $a$ and $b$ are constants, and $C_1$ is a turbulent natural convection coefficient.

**Table 10** MoWiTT Coefficients (Yazdanian and Klems 1994; Booten et al. 2012)

| Wind Direction | $C_1$ (W m$^{-2}$K$^{-(4/3)}$) | $a$ (W m$^{-2}$K (m s$^{-1}$)$^b$) | $b$ - |
|---|---|---|---|
| Windward | 0.84 | 3.26 | 0.89 |
| Leeward | 0.84 | 3.55 | 0.617 |

The DOE-2 method employs the following method to calculate the convection coefficient:

$$h = h_{natural} + R_f(h_c - h_{natural}) \tag{52},$$

where, $R_f$ is the surface roughness multiplier (Table 8) and $h_{natural}$ is the natural convection coefficient that is calculated from Eq. (48), (49), and (50).

# Appendix D Moisture Model

Table 11 Parameters required to run the moisture model and their default values for the moisture model validation case in Sect. 3.5 (based on Table 1 of Matthews, 2006)

| Group | Parameter | Units | Value |
|---|---|---|---|
| Solid fuel | Layer depth | m | 0.02 |
| | Bulk density | kg m$^{-3}$ | 62 |
| | Solid fuel density | kg m$^{-3}$ | 550 |
| | Layer surface area to volume ratio | m$^{-1}$ | 3000 |
| | Characteristic length | m | 3000 |
| Fuel moisture | Nelson (1984) model parameter A | | 5.2 |
| | Nelson (1984) model parameter B | | -19 |
| | Solid fuel surface conductance | m s$^{-1}$ | 0.0006 |
| | Saturation moisture content | kg kg$^{-1}$ | 1.4 |
| | Liquid water absorption parameter $L_a$ | kg kg$^{-1}$ | 0.23 |
| | Liquid water absorption parameter $L_b$ | kg kg$^{-1}$ | -1.63 |
| Radiation | Solid fuel albedo | | 0.27 |
| | Attenuation coefficient | | 1.363 |
| Conduction | Heat conductivity as a function of moisture content $K_{C,slope}$ | W m$^{-1}$ K$^{-1}$ | 0.2 |
| | Heat conductivity as a function of moisture content $K_{C,intercept}$ | W m$^{-1}$ K$^{-1}$ | 0.14 |
| | Solid fuel to water heat conductivity $K_{H,f,w}$ | W m$^{-2}$ K$^{-1}$ | 700 |
| Drainage | Rainfall storage capacity | kg kg$^{-1}$ | 1.153 |
| | Drainage coefficient | s$^{-1}$ | 0.00003 |
| Vertical Mixing | Diffusivity at top of fuel layer $D_{T0,a}$ | m$^2$ s$^{-1}$ | 0.00002 |
| | Diffusivity at top of fuel layer $D_{T0,b}$ | s m$^{-1}$ | 2.6 |
| | Attenuation coefficient $\chi_a$ | s m$^{-1}$ | 2.08 |
| | Attenuation coefficient $\chi_b$ | s m$^{-1}$ | 2.38 |
| Soil | Soil albedo | | 0.2 |
| | Soil fuel capacity | m$^2$ m$^{-3}$ | 0.3 |
| Top Boundary | Aerodynamic roughness length | m | 0.01 |
| | Screen height | m | 2 |
| Constants | Thermal diffusivity of dry air | m$^2$ s$^{-1}$ | 2.08×10$^{-5}$ |
| | Diffusion coefficient for water vapor in air | m$^2$ s$^{-1}$ | 2.34×10$^{-5}$ |
| | Latent heat of vaporization of water | J kg$^{-1}$ | 2.45×10$^6$ |
| | Specific heat of air | J kg$^{-1}$ K$^{-1}$ | 1004.5 |
| | Von Karman constant | | 0.4 |

# Data Availability

The data that support the findings of this study are available from the corresponding author upon reasonable request.